\documentclass[osajnl,twocolumn,showpacs,superscriptaddress,10pt]{revtex4-1} 
\usepackage{amsmath,amssymb,graphicx}
\begin{document}

\title{Stratified dispersive model for material characterization using THz time-domain spectroscopy}

\author{J.L.M. van Mechelen}\email{Corresponding author: dook.vanmechelen@ch.abb.com}
\affiliation{ABB Corporate Research, Segelhofstrasse 1K, 5405 Baden-D\"attwil, Switzerland}
\author{A.B. Kuzmenko}
\affiliation{D\'epartement de Physique de la Mati\`ere Condens\'ee, Universit\'e de Gen\`eve, Gen\`eve, Switzerland}

\author{H. Merbold}
\affiliation{ABB Corporate Research, Segelhofstrasse 1K, 5405 Baden-D\"attwil, Switzerland}

\begin{abstract} We propose a novel THz material analysis approach which provides highly accurate material parameters and can be used for industrial quality control. The method treats the inspected material within its environment locally as a stratified system and describes the light-matter interaction of each layer in a realistic way. The approach is illustrated in the time- and frequency-domain for two potential fields of implementation of THz technology: quality control of (coated) paper sheets and car paint multilayers, both measured in humid air.
\end{abstract}


\maketitle 

The last decade has seen a significant development in the maturity of THz technology which is paving the road towards industrial applications. THz technology has the potential to become a real differentiator given its unusual properties ranging from the sensitivity to depth information to providing complex optical material functions while being innocuous for human tissue. A field of application which would substantially benefit from these aspects is material quality control in automatized processes for detecting e.g. in-depth failures. Recently, THz spectroscopy based quality control has been explored on materials such as paper~\cite{White2011}, food, plastics, semiconductors, and biological tissues.\cite{Tonouchi2007} Common methods to obtain the material properties range from time-domain peak subtraction~\cite{White2011,Yasui2005}  to inversion of the transfer function~\cite{Duvillaret1999,Dorney2001}. Both methods require several internal reflections and typically apply to single layered materials which are assumed to have low absorption and little dispersion. Recently, a time-domain fitting procedure has been proposed for multilayers,\cite{Beigang2013} although it requires prior knowledge of the optical material properties and only provides the thicknesses. Despite the fact that these methods may give satisfactory material parameters in some situations, they break down or lack accuracy when applied to industrial quality control where the inspected material and its environment form a complex structure with properties that may change over time.

In this Letter we describe a novel THz material analysis approach allowing for high precision material parameter determination of complex structures, consisting of (i) a stratified dispersive model, (ii) an appropriate measurement configuration and (iii) a time-domain based fitting procedure. The underlying concept is to model the measurement configuration as a stratified system where for each layer the physical processes that occur upon the light-matter interaction are described in a realistic way. Subsequently, the light propagation through the multilayer system is calculated and fitted to the experimental data in the time-domain, thereby optimizing all material parameters together. The method is demonstrated for two prominent industrial examples, inspection of (coated) paper sheets and car paint multilayers. It is in particular shown that the method is applicable in ambient air with variable humidity and irrespective of the reference position in reflection geometry, two of the major obstacles encountered in today's THz analysis.

\begin{figure}[b]
\centerline{\includegraphics[width=0.925\columnwidth]{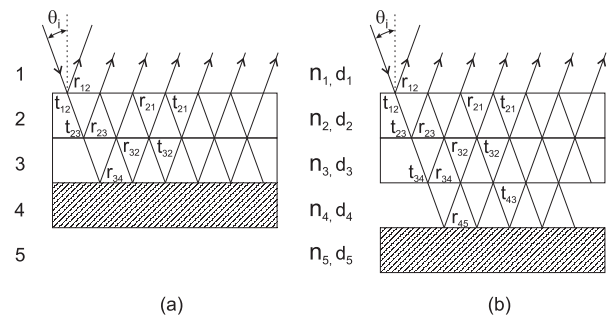}}
\caption{Schematic representation of a light ray incident at angle $\theta_i$ interacting with a double layer structure (a) on a reflector and (b) suspended above a reflector, in ambient air.}
\label{fig1}
\end{figure}

The measurement configuration in common optical setups consists of the material under investigation positioned in a certain environment, e.g. ambient air. In industrial circumstances, a reflection geometry is often preferable to a transmission arrangement due to more convenient accessibility of the sample. The overall strength of the reflected signal can be increased by positioning a (metal) reflector behind the material~\cite{White2011}. A central principle of the proposed analysis approach is to locally consider the measurement configuration, composed of the inspected material, the surrounding environment and the reflector, as a stratified system. The advantage is that the light propagation through this system can be analytically described by the Fresnel equations. For the case of a bilayer material system shown in Fig.~\ref{fig1}a with the angle of incidence $\theta_i=0$, the reflected electric field $E_r$ can be calculated using the incident electric field $E_{r,0}$ and the transfer function of the entire multilayer structure $\mathcal{T}$ through

\begin{equation}\label{fresnel}
\begin{split}
E_{r}(\omega)=&\ \mathcal{T}(\omega)E_{r,0}(\omega)\\
 \mathcal{T}(\omega) = &\ r_{12}+t_{12}r_{23}t_{21}e^{-i2\beta_2}+t_{12}r_{23}r_{21}r_{23}t_{21}e^{-i4\beta_2}\\
  &+t_{12}t_{23}r_{34}t_{32}t_{21}e^{-i2(\beta_2+\beta_3)}+\ldots
\end{split}
\end{equation}
where $\beta_k=\omega n_k d_k/c$ is the phase shift accumulated in layer $k$, $\omega$ is the frequency of the radiation, $n_k$ is the complex index of refraction of layer $k$, $d_k$ is the thickness of layer $k$, $c$ is the speed of light in vacuum, and the transmission $t_{ij}$ and reflection $r_{ij}$ coefficients are
\begin{equation}\label{fresnelcoeff}
 t_{ij}=\frac{2n_i}{n_i+n_j},\quad\qquad r_{ij}=\frac{n_i-n_j}{n_i+n_j}.
\end{equation}
The full expression of $\mathcal{T}(\omega)$ as well as the case of oblique incidence can be found in standard optics literature~\cite{Anders}.

The stratified model (Eqs.~\ref{fresnel}-\ref{fresnelcoeff}) contains $d_k$ and $n_k$ of each individual layer, opposite to simple analysis methods which only consider $d$ and $n$ of the material layer, and furthermore assume $n$ being real and frequency independent. We assume that $d_k$ and $n_k$ are not known, and aim at a realistic description of $n_k$ by addressing the individual physical processes in each layer which results in $n_k$ being complex and dispersive. Although academic studies have shown that many exotic materials, such as semi- or superconductors, are characterized by diverse light induced processes, materials encountered in industrial inspection applications are often much simpler. In such common materials the interaction with THz radiation is typically limited to lattice vibrations and (free and/or collective) electron oscillations which can often be well described by Lorentzian line shapes in $n_k(\omega)$. However, the analysis method is general in a sense that interactions may also be described by other line shapes such as Gaussian, Fano or Tauc-Lorentz. An appropriate model for $n_k(\omega)$ is thus a summation of oscillators, each representing a specific excitation. For the case of a series of Lorentzians, this model is commonly known as the Drude-Lorentz parameterization and the dielectric function $\epsilon(\omega)=n(\omega)^2$ of each layer $k$ is given by

\begin{equation}\label{DrudeLorentz}
  \epsilon(\omega)=\epsilon_{\infty}+\sum_{\ell=1}^m \frac{\omega_{p,\ell}^2}{\omega_{0,\ell}^2-\omega^2-i\gamma_\ell\omega}
\end{equation}
where $\epsilon_{\infty}$ is the high frequency limit of $\epsilon(\omega)$, $\omega_{p,\ell}$ the plasma frequency, $\omega_{0,\ell}$ the characteristic frequency and $\gamma_\ell$ the relaxation rate of excitation $\ell$ in layer $k$. For a measurement configuration as shown in Fig.~\ref{fig1}b the two material layers 2 and 3 are parameterized in this way. The thin air layers 1 and 4, on the other hand, can simply be modeled by fixing $n_{\text{air}}=1+0i$ through setting $\epsilon_{\infty,\text{air}}=1$ without further Drude-Lorentz parameters. For the metallic reflector, layer 5, a single fixed Drude oscillator is used which accounts for the reflectivity of the metal in the THz range. In this way $n_k$ describes the interaction with each layer $k$ in a realistic way. Substitution of $n_k$ into Eqs.~\ref{fresnel}-\ref{fresnelcoeff} now fully describes $E_r$ of the examined material in its measurement configuration.

From an experimental perspective, quantitative information of a material can be obtained by performing both a measurement of the sample structure and of a known reference, leading to a data set consisting of $|E(\omega)|^2$ and $|E_0(\omega)|^2$, respectively. The goal of the analysis is to obtain $d_k$ and $n_k(\omega)$ of the probed material layer(s). Model free analytical inversion of Eq.~\ref{fresnel} is often applied for single or double layers~\cite{Beigang2013}, but structures as shown in Fig.~\ref{fig1} are from a practical point of view too complicated for this approach. The commonly used alternative is fitting $\mathcal{T}$ of Eq.~\ref{fresnel} to the reflectivity $|r(\omega)|^2=|E_r(\omega)/E_{r,0}(\omega)|^2$. However, multiple internal reflections occurring in structures as shown in Fig.~\ref{fig1} make $|r(\omega)|^2$ rarely easy to understand. The more comprehensive data set is composed by the time-domain functions $E_r(t)$ and $E_{r,0}(t)$ which often clearly reveal the partial reflections from the various interfaces. Central to the proposed analysis method is to perform a time-domain fitting procedure of $E_r(t)$ to the experimentally determined $E_r^{\text{exp}}(t)$ using a least squares algorithm. Hereto, the transfer function $\mathcal{T}$ of Eq.~\ref{fresnel} has to be Fourier transformed to the time domain and subsequently convolved with $E_{r,0}(t)$. It can now be seen that the purpose of having a thin air layer 1 in Fig.~\ref{fig1}, modeled as $n_{\text{air}}=1$, is to account for the possible mismatch between the sample and reference position. Namely, the fitting parameter $d_{\text{air}}$  corresponds uniquely to a temporal displacement of $E_r(t)$,  and thus to the relative position of sample and reference. Another major advantage of the approach is that ambient humidity present in the reference $E_{r,0}(t)$ is inherently (see Eq.~\ref{fresnel}) also present in $E_{r}(t)$ which allows analysis in ambient air without loss of accuracy.

In the following we will illustrate the robustness of the analysis method by giving two experimental examples. The first one is the quality control of paper for which the industry requires sensitivity to a variety of parameters on which it imposes strict accuracy standards. We apply the proposed method to obtain the thickness and the ash (filler) concentration of various kinds of paper sheets.

\begin{figure}[htbp]
\centerline{\includegraphics[width=0.85\columnwidth]{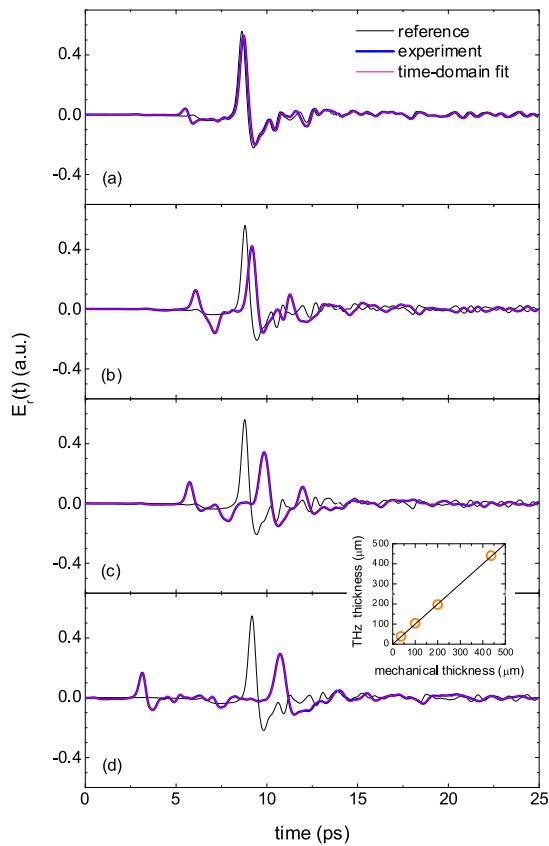}}
\caption{$E_r^{\text{exp}}(t)$ of (a-c) uncoated and (d) coated paper sheets, $E_{r,0}^{\text{exp}}(t)$ of a copper reflector, and fits $E_r(t)$, using s-polarized incident radiation, $\theta_i=13^\circ$, $26\pm2\,^\circ$C and $22\pm2$ \% relative humidity (RH). (a) tissue paper, (b) 100 g/m$^2$ copy paper, (c) 200 g/m$^2$ copy paper, (d) coated board. The inset shows the total thickness of the samples (a-d) obtained from the fitting procedure as compared to mechanical caliper values.}
\label{fig2}
\end{figure}

\begin{table}[b]
\begin{center}\caption{\label{table1}THz thickness values of various paper sheets as compared to values from rated mechanical techniques (in $\mu$m).}
\begin{tabular}{l|c|c}
\hline
paper sample & \ rated thickness\  & \ THz thickness\ \\ \hline
tissue & 37 &  38.7 \\
100 g/m$^2$ copy paper & $106\pm4$ & 104.0 \\
200 g/m$^2$ copy paper & $200\pm6$ & 197.4 \\
single sided coated board & $437\pm4$ & 40.4 + 400.4\\ \hline
\end{tabular}
\end{center}
\end{table}

Fig.~\ref{fig2} shows the experimental $E_r^{\text{exp}}(t)$ of three uncoated and one coated papers sheet with varying thickness, and $E_{r,0}^{\text{exp}}(t)$ of an optically polished copper reference measured by THz time-domain spectroscopy (TAS\,7500, Advantest Inc.) in the range 0.1-4 THz, recorded with 512 averages at 125 Hz. The measurement configuration consists of air-(coating)-paper-air-copper (see Fig.~\ref{fig1}b) and can thus be modeled as a five (four) layer system for (un)coated paper. In order to perform a time-domain fit to $E_r^{\text{exp}}(t)$, we need to determine $n_k$ for each layer $k$ of the system. Setting $n_k$ of air and copper as discussed before, the question arises how the interaction of THz radiation with paper can be described. The usage of too many oscillators could provide a satisfactory fit, but their parameters may not have any physical meaning. It turns out that the fibrous structure of paper slightly modifies $n$ from being frequency independent. Paper may further contain so-called ash which is a mineral that fills the spaces between the fibers and  often features a characteristic phonon absorption in the THz range. This means that paper can typically be modeled with one or two oscillators, one for the fibrous structure with $\omega_0\approx 20-200$ THz and in case it also contains ash, like copy paper, a second one with $\omega_0<10$\,THz. Tissue paper just contains fibers, whereas the coating of the board consists of only ash. Based on the experimental data, we choose the oscillators to be Lorentzians and parameterize them as given by Eq.~\ref{DrudeLorentz}. At this stage, we have fully described the optical properties of the entire multilayer configuration at THz frequencies, which allows to calculate $E_r(t)$ using Eq.~\ref{fresnel} and fit it to $E_r^{\text{exp}}(t)$. The fitting parameters are $d_k$, and $\epsilon_\infty,\omega_0,\omega_p,\gamma$  which model the paper sheet and coating layer. The result of the fitting procedure is shown in Fig.~\ref{fig2} and the obtained thicknesses are shown in Table~\ref{table1} where they are  compared to rated values measured by standard mechanical techniques (see also inset of Fig.~\ref{fig2}). The excellent fit results together with the close match of the thickness values demonstrates the robustness of the approach. Moreover, the fit accurately describes the ambient moisture present in $E_r^{\text{exp}}(t)$ and accounts for the mismatch of the reference position with respect to the sample (here around 0.5 mm), both without compromising the accuracy. Note that although $E_r^{\text{exp}}(t)$ has been recorded in 4\,s, industrial paper quality control typically allows much shorter integration times.  This will add noise on the amplitude of $E_r^{\text{exp}}(t)$, but affects less the temporal information, which is of main importance for the thickness determination using the proposed method.


\begin{figure}[b]
\centerline{\includegraphics[width=0.7\columnwidth]{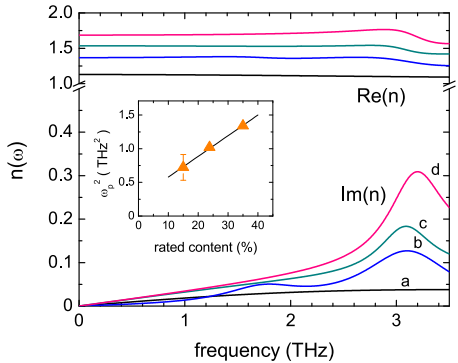}}
\caption{Real and imaginary part of $n(\omega)$ of (a) tissue paper, (b-c) copy paper, (d) magazine paper. The inset shows $\omega_p^2$ vs.~the rated CaCO$_3$ content. The error bar gives the spread in $\omega_p^2 $ of 5 sheets of different basis weight of the same grade. Note that as compared to samples (c) and (d), copy paper (b) requires an additional oscillator around 1.8\,THz.}
\label{fig3}
\end{figure}

The filler concentration, on the other hand, can be determined by the strength of the ash's phonon absorption as manifested in $n_k(\omega)$ of the paper sheet. Fig.~\ref{fig3} shows the real and imaginary part of $n(\omega)$ of several paper samples containing CaCO$_3$ ash, obtained from a time-domain fitting analysis as outlined above. For comparison, tissue paper which contains no ash is also shown. For all ash-containing samples Im\,$n(\omega)$ shows a strong increase around 3 THz which corresponds well to the literature value of the CaCO$_3$ lattice vibration~\cite{Mizuno}. The spectral weight of the absorption is proportional to $\omega_p^2$ which is directly obtained from the analysis procedure. The correlation of $\omega_p^2$ with the rated CaCO$_3$ content (as stated by the paper manufacturer) suggests a linear behavior in the measured ash range (see inset of Fig.~\ref{fig3}). In addition to the thickness, the analysis method can thus also provide information on the consistency of paper.

\begin{table}
\begin{center}\caption{\label{table2}THz thickness values (in $\mu$m) of a triple paint layer on steel and silicon (see text) compared to rated techniques.}
\begin{tabular}{l|c|c|c|c}
\hline
sample &\ THz individual\ &\  THz total\  &  mechanical  &  magnetic \\ \hline
steel & 42.5, 22.3, 31.9 & 96.7 &  $98\pm1$ & $97\pm5$\\
silicon & 45.7, 17.0, 38.3 & 101.0 & $100\pm3$ & $101\pm3$\\ \hline
\end{tabular}
\end{center}
\end{table}

In a second example, we have applied the analysis method to automotive paint layers, another widely suggested industrial application of THz technology~\cite{Yasui2005}. Different kinds of Glasurit BASF car paints were used in order to make single, double and triple layer structures on a substrate as employed in the automotive industry. Although up to now we have solely dealt with reflection geometries, in certain cases industrial inspection may also require transmission geometries, for which the stratified dispersive model (cf.~Eqs.~\ref{fresnel}-\ref{fresnelcoeff}) can be straightforwardly extended. In order to demonstrate both configurations, we have used substrates of both steel and silicon for reflection and transmission measurements, respectively.

\begin{figure}[htbp]
\centerline{\includegraphics[width=0.85\columnwidth]{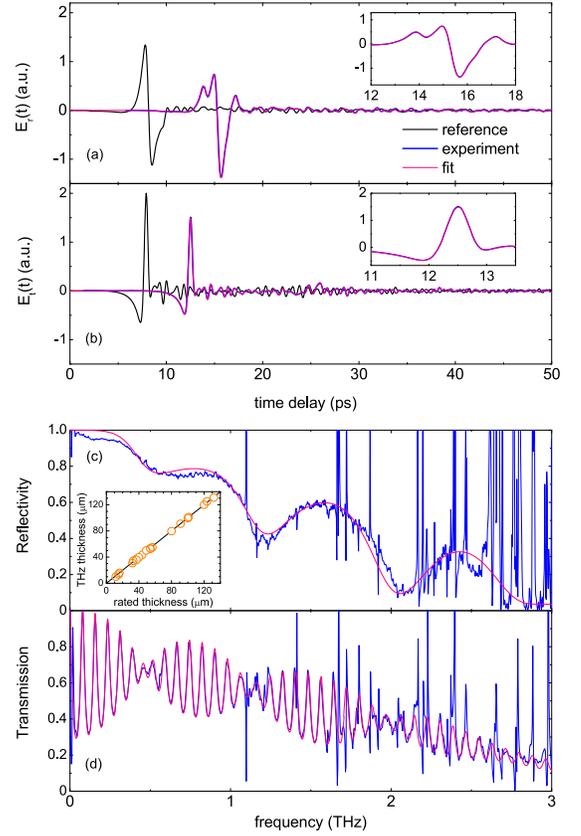}}
\caption{(a) $E_r^{\text{exp}}(t)$ and $E_{r,0}^{\text{exp}}(t)$, (b) $E_t^{\text{exp}}(t)$ and $E_{t,0}^{\text{exp}}(t)$, (c) the corresponding $|r^{\text{exp}}(\omega)|^2$ and (d) $|t^{\text{exp}}(\omega)|^2$ of a triple paint layer on (a,c) steel and (b,d) silicon, recorded using 200 averages at 30 Hz, at $26\pm1\,^\circ$C and $55\pm1$ \%\,RH, and the best fit results based on the method. The inset of panel (c) shows for a larger set of samples the determined total paint layer thickness as compared to values from rated techniques.}\label{fig4}
\end{figure}

Fig.~\ref{fig4}a,b shows $E_r^{\text{exp}}(t)$ and $E_t^{\text{exp}}(t)$ of a triple paint layer consisting of white primer, blue waterborne base coat and clear coat on steel and silicon, respectively, as measured by THz time-domain spectroscopy (TPI 1000, TeraView Ltd.) in the range 0.03-3 THz. As compared to Fig.~\ref{fig2}, $E_r^{\text{exp}}(t)$ seems less complicated and appears as single cycle pulses with little structure. This is, however, deceptive since the small layer thicknesses and the absence of an air gap between paint and substrate leads to an overlap of the multiple internal reflections. Fig.~\ref{fig4}c,d shows the corresponding $|r^{\text{exp}}(\omega)|^2$ and $|t^{\text{exp}}(\omega)|^2$ in the frequency-domain. In this example, we have applied the proposed analysis method in both the time- and frequency-domain simultaneously (see Fig.~\ref{fig4}). This enhances the accuracy of the fit parameters in case $|r(\omega)|^2$ and/or $|t(\omega)|^2$ is strongly frequency dependent, as in the present example. The thicknesses resulting from the fit are shown in Table~\ref{table2} for each individual layer as well as for the total stack, and are compared to rated values from
 mechanical caliper and magnetic induction techniques. The inset of Fig.~\ref{fig4}c shows the total thickness values of a larger set of single and multilayer samples versus the rated values. In all cases the comparison is very close which proves the accuracy of the analysis method.

In conclusion, we have proposed a widely applicable method to perform and analyze THz measurements of complex structures which is able to provide accurate material parameters. The approach treats the inspected material and its environment as a stratified system and for each layer parameterizes the material properties using a set of oscillators. The method has been demonstrated for paper and paint multilayer samples in both the time- and frequency-domain, for reflection and transmission geometries in humid ambient air.

We would like to thank Dirk van der Marel for the use of the TeraView spectrometer and Xun Gu for a critical reading of the manuscript.

\end{document}